%% file: QuboGSLinEqSolver.tex
\documentclass[aps,pra,twocolumn,superscriptaddress, floatfix]{revtex4-2}
\usepackage{graphicx}
\usepackage{amsmath}
\usepackage{amssymb}
\usepackage{color}
\usepackage{physics}
\usepackage{verbatim}
\usepackage{graphicx}
\usepackage{bm}

\usepackage{amsmath}
\usepackage{physics}
\usepackage{qcircuit}
\usepackage{listings}
\usepackage{xcolor}
\usepackage{enumitem}
\definecolor{codegreen}{rgb}{0,0.6,0}
\definecolor{codegray}{rgb}{0.5,0.5,0.5}
\definecolor{codepurple}{rgb}{0.58,0,0.82}
\definecolor{backcolour}{rgb}{0.95,0.95,0.92}
 
\lstdefinestyle{mystyle}{
    backgroundcolor=\color{backcolour},   
    commentstyle=\color{codegreen},
    keywordstyle=\color{magenta},
    numberstyle=\tiny\color{codegray},
    stringstyle=\color{codepurple},
    basicstyle=\ttfamily\footnotesize,
    breakatwhitespace=false,         
    breaklines=true,                 
    captionpos=b,                    
    keepspaces=true,                 
    numbers=left,                    
    numbersep=5pt,                  
    showspaces=false,                
    showstringspaces=false,
    showtabs=false,                  
    tabsize=2
}
\usepackage{float}   
\begin{document}

\input{abstract}
\input{preamble}
\input{introduction}

\input{methods}
\input{results}

\input{acknowledgments}
\input{conclusions}

\bibliography{apssamp}
\end{document}

%% file: abstract.tex
\begin{abstract}
The numerical solution of partial differential equations by discretization techniques is ubiquitous in computational physics. In this work we benchmark this approach in the quantum realm by solving the heat equation for a square plate subject to fixed temperatures at the edges and random heat sources and sinks within the domain. The hybrid classical-quantum approach consists in the solution on a quantum computer of the coupled linear system of equations that result from the discretization step. Owing to the limitations in the number of qubits and their connectivity, we use the Gauss-Seidel method to divide the full system of linear equations into subsystems, which are solved iteratively in block fashion. Each of the linear subsystems were solved using 2000Q and Advantage quantum computers developed by D-Wave Systems Inc. By comparing classical numerical and quantum solutions, we observe that the errors and chain break fraction are, on average, greater on the 2000Q system. Unlike the classical Gauss-Seidel method, the errors of the quantum solutions level off after a few iterations of our algorithm. This is partly a result of the span of the real number line available from the mapping of the chosen size of the set of qubit states. We verified this by using techniques to progressively shrink the range mapped by the set of qubit states at each iteration (increasing floating-point accuracy). As a result, no leveling off is observed. However, an increase in qubits does not translate to an overall lower error. This is believed to be indicative of the increasing length of chains required for the mapping to real numbers and the ensuing limitations of hardware.
\end{abstract}

%% file: preamble.tex

\title{A hybrid classical-quantum approach to solve the heat equation using quantum annealers}

\author{Giovani G. Pollachini}
\affiliation{
 QuanBy Computa\c c\~ao Qu\^antica, Florianópolis, Brazil.
}%

\author{Juan P. L. C. Salazar}
\affiliation{Engenharia Aeroespacial, Universidade Federal de Santa Catarina, Joinville, SC, 89.219-600, Brazil}

\author{Caio B. D. Góes}
\affiliation{Departamento de F\' \i sica, Universidade Federal de Santa Catarina, Florian\'opolis, SC, 88040-900, Brazil}

\author{Thiago O. Maciel}
\affiliation{Departamento de F\' \i sica, Universidade Federal de Santa Catarina, Florian\'opolis, SC, 88040-900, Brazil}

\author{Eduardo I. Duzzioni} 

\affiliation{
 QuanBy Computa\c c\~ao Qu\^antica, Florianópolis, Brazil.
}
\affiliation{Departamento de F\' \i sica, Universidade Federal de Santa Catarina, Florian\'opolis, SC, 88040-900, Brazil}

\email{duzzioni@gmail.com}

\date{\today}
\maketitle

%% file: introduction.tex
\section{\label{sec:introduction}Introduction}

In fields of computational physics, the mathematical description of the problem of interest relies on the solution of partial differential equations restricted to boundary conditions and initial conditions. Most of the techniques are based on discretization methods, where the physical domain is divided into cells or volumes. When implicit methods are chosen, the discrete set of coupled equations to be solved are cast as a system of linear equations of the form $A\mathbf{x}=\mathbf{b}$, where $A$ is generally a $N\times N$ sparse matrix and the vector $\mathbf{b}$ is known, while the vector $\mathbf{x}$ is the sought solution to the problem. The largest fraction of computational time is spent on the solution of the linear systems.

After the advent of the Harrow, Hassidim, and Lloyd (HHL) algorithm \citep{harrow2009quantum} to solve the linear quantum problem, subsequent work has used it as subroutine to solve ordinary or partial differential equations \cite{leyton2008quantum,clader2013preconditioned,cao2013quantum,berry2014high,montanaro2016quantum,berry2017quantum,costa2019quantum,wang2020quantum}. However, the solutions for the differential equations were presented in form of a vector state, restricted to the calculation of global properties of the system of interest, otherwise the cost to obtain each amplitude of the quantum state would render the quantum approach too expensive. \citet{clader2013preconditioned} argue that an exponential speedup can be achieved by using the HHL algorithm to solve the electromagnetic scattering cross-section problem via the finite element method. However, \citet{montanaro2016quantum} showed that the relation between the size of the system of linear equations and the accuracy of the solution was missed, which implies that the speedup can be reduced or even eliminated. A thorough analysis of the resources needed to solve the electromagnetic scattering cross-section of a 2D target was made by \citet{scherer2017concrete}, concluding that the number of resources must be reduced by many orders of magnitude for the algorithm to become practical. There are other issues concerning quantum state preparation and the necessity of an oracle to encode the coefficients of the matrix $A$. For reasons like these, it is not yet clear if exponential speedups have been achieved by quantum algorithms using the linear quantum problem as a subroutine. Recently, \citet{linden2020quantum} showed that HHL-like algorithms used to solve the heat equation are never faster than the best classical algorithms. However, a quantum alternative which uses amplitude amplification can accelerate the process, achieving at most a quadratic speedup. Although this a very encouraging result, the gate model of quantum computing needs advance to approach large problems. Following this reasoning, quantum annealers can be more attractive at the current stage of quantum computing due to their large number of qubits and increase connectivity. For instance, a graph-coloring-based methodology to solve differential equations through quantum annealers was proposed in \cite{srivastava2019box}.

An alternative method for solving the linear system of equations by quantum annealing was presented in Refs. \cite{o2016toq} and \citet{rogers2019floating}. An analysis of the resources employed for this task was made in Ref. \cite{borle2019analyzing}, showing that in some circumstances it is possible to have advantage when compared to traditional algorithms to solve least squares problems. The linear system is encoded in a QUBO (\textit{quadratic unconstrained binary optimization}) problem \cite{glover2018tutorial} through a well-defined protocol, whose ground state is the solution to our problem of interest. As opposed to a HHL-type algorithm, the final solution of the linear system of equations is a classical state codifying the vector $\mathbf{x}$. In this work we solve the steady-state 2D heat equation by a hybrid classical-quantum algorithm. Applying the finite difference method to the partial differential equation yields a system of coupled linear equations. In order to enable the solution of larger linear systems of equations, we use the block Gauss-Seidel method and solve the full system iteratively, where the smaller subsystems of equations are solved in the quantum processing units (QPU). The quantum algorithm is implemented in the 2000Q and Advantage systems of D-Wave Systems Inc. 

The paper is divided as follows. In \S\ref{sec:methods} we introduce the D-Wave quantum annealing methodology, review the approach of \citet{rogers2019floating} to solve the linear system of equations, present the block Gauss-Seidel method, and elaborate about quantum resources required to implement the algorithm. The results are presented in \S\ref{sec:results} and are followed by conclusions and perspectives in \S\ref{sec:conclusion}.

%% file: methods.tex
\section{\label{sec:methods}Methods}

As described above, the basic idea of our hybrid approach is to replace a classical subroutine to solve the system of linear equations that result from the discretization of partial differential equations by a quantum algorithm. In the proposed method the system of linear equations is solved using quantum annealing, where the final Hamiltonian encodes a QUBO problem \citep{rogers2019floating}. In the following we briefly review the quantum annealing followed by the QUBO problem. 

\subsection{D-Wave quantum annealing}
D-Wave quantum annealing is based on the adiabatic quantum evolution of the transverse Ising Hamiltonian with the system in thermal equilibrium at a very low temperature \citep{kadowaki1998quantum}. The goal of the computation is to minimize an objective function (energy) for which the solution to the desired problem is approximately encoded in its ground state \cite{apolloni1989quantum}. In order to achieve the ground state of the system with a high success rate, the quantum fluctuations must be dominant over thermal fluctuations \citep{kadowaki1998quantum}.      


The idea behind this model is the preparation of the system as the ground state of an easy-to-prepare Hamiltonian, for instance ($\hbar\omega=1$),
\begin{align}
H(0) = -\sum_{i=1}^n \sigma_x^i,
\end{align}
where $\sigma_x^i$ is the Pauli matrix in the $x$ direction for the $i$-particle and its ground state is $|\psi(0)\rangle=|+\rangle^{\otimes n}$, which contains the superposition of all states in the computational basis equally distributed. By choosing the final Hamiltonian $H(1)$ as the operator whose ground state contains the solution to the desired problem, it is possible to interpolate between these two Hamiltonians with a schedule function $f(s)$, subject to $f(0)=0$ and $f(1)=1$, resulting in,
\begin{align}
H(s)=\left(1-f(s)  \right)H(0) + f(s) H(1), 
\end{align}
where $s=t/T$ ($0 \le s \le 1$) is the dimensionless time, $t$ is the current time and $T$ is the total time of the computation. The final Hamiltonian implemented by D-Wave is the Ising Hamiltonian,
\begin{align}\label{eq:ising_hamiltonian}
    H(1)= H_{Ising} = \sum_{i} \alpha_{i} \sigma_z^{i} + \sum_{i,j} \beta_{ij} \sigma_z^{i} \sigma_z^{j}  \ ,
\end{align}
with the parameters $\alpha_{i}$ and $\beta_{ij}$ being chosen to encode the solution to the problem and $\sigma_z^{i}$ is the Pauli matrix in the $z$ direction for the particle $i$.

According to the adiabatic theorem \cite{jansen2007bounds}, if the evolution is performed adiabatically, then the ground state of the problem Hamiltonian $H(1)$ is found with high probability, provided that the evolution time $T$ from the initial Hamiltonian $H(0)$ to $H(1)$ is proportional to $g_{min}^{-2}$, where $g_{min}=E_1-E_0$ is the minimum energy gap between the ground and first excited states \cite{jansen2007bounds,albash2018adiabatic}. It is important to observe that a suitable choice for the schedule function may result in totally different evolution times \cite{roland2002quantum,albash2018adiabatic,an2019quantum}.
Therefore, the final Hamiltonian $H(1)$ must contain the solution to the linear system of equations \citep{rogers2019floating}. Next, we show how to build $H(1)$, thus casting the system of linear equations into a QUBO problem.

\subsection{QUBO algorithm to solve linear systems of equations}

\citet{o2016toq} and \citet{rogers2019floating} showed how to map the problem of solving a linear system of equations into a QUBO problem \citep{glover2018tutorial}. Also, in Ref. \cite{chang2019quantum}, the linear system of equations can be obtained as a particular case of a nonlinear system of equations. Given a linear system $A\mathbf{x} = \mathbf{b}$, the solution $\mathbf{x}$ can be found by minimizing the function
\begin{align}\label{eq:ls_as_min_problem}
    H(\mathbf{x}) = (A\mathbf{x} - \mathbf{b})^{\dag} (A\mathbf{x} - \mathbf{b}) \ .
\end{align}
We can approximate each component $x_i$ ($i=1\dots N$) of the vector $\mathbf{x}$ 
by its binary representation in a given interval $[-d_i,2c_i-d_i)$, with $R$ bits, as:
\begin{align}\label{eq:xi_binary}
    x_i = c_i\sum_{r=0}^{R-1}q_r^i 2^{-r} - d_i  \ , 
\end{align}
where each $q_r^i \in \{ 0,1 \}$ is the $r$-th bit in the binary representation of $x_i$. If $d_i>0$ and $c_i>d_i/2$, the domain of $x_i$ contains positive and negative real values. This can easily be generalized to complex variables \citep{rogers2019floating}. After substituting (\ref{eq:xi_binary}) into (\ref{eq:ls_as_min_problem}) and dropping the positive constant term, the equation becomes,
\begin{align}\label{eq:qubo_ls_problem}
    H(q) = \sum_{r=0}^{R-1} \sum_{i=0}^{N-1} a_r^i q_r^i + \sum_{r,s=0}^{R-1} \sum_{i,j=0}^{N-1} b_{rs}^{ij} q_r^i q_s^j  \ ,
\end{align}
where the coefficients are given by,
 \begin{align}
    a_r^i &= -2 \left( \sum_{j,k=0}^{N-1} A_{ki} A_{kj} c_i d_j + \sum_{j=0}^{N-1} A_{ji} c_i b_j \right) 2^{-r} \label{eq:qubo_coefficients1} \ ,   \\
      b_{rs}^{ij} &= \left(\sum_{k=0}^{N-1} A_{ki} A_{kj} c_i c_j \right) 2^{-(r+s)} \label{eq:qubo_coefficients2}.
\end{align} 
Here, $A_{ij}$ are coefficients of the matrix $A$ and $b_i$ are components of the vector $\mathbf{b}$. It is important not to confuse $b_i$ with $b^{ij}_{rs}$.



In quantum annealers, the physical qubits are indexed by a 1-dimensional linear index $\ell$ (see Eq. (\ref{eq:ising_hamiltonian})), while the logical qubits are indexed by 2-dimensional indices $i$ and $r$, where $i = 0,1,...,N-1$ and $r=0,1,...,R-1$, as shown in Eq. (\ref{eq:qubo_ls_problem}). The mapping between the logical an physical indices is given by,
\begin{align}
    \ell (i,r)&=i.R+r, \,\, \ell = 0,...,NR-1, \,\,
\end{align}
such that the inverse mapping is,
\begin{align}
    i_\ell &= \lfloor \ell/R \rfloor \ , \nonumber\\
    r_{\ell}&= \ell \mod R.
\end{align}
The mapping from $q_{\ell} \rightarrow \sigma_{\ell}$ is $\sigma_{\ell}=2q_{\ell}-1$ with $\sigma_{\ell}=\{+1,-1\}$.

Finally, the quantization of the QUBO problem is made through $\hat{q}_{\ell}|q\rangle =q_{\ell}|q\rangle$, where the operators are such that $\hat{q}^2_{\ell}=\hat{q}_{\ell}$, with eigenvalues $q_{\ell} \in \{0,1\}$ and $|q\rangle=|q_0\rangle\otimes ...\otimes |q_{NR-1}\rangle$. 
In order to solve the system of linear equations, the coefficients $a^i_r \rightarrow a_{\ell}$ and $b^{ij}_{rs} \rightarrow b_{\ell \kappa}$, described in Eqs. (\ref{eq:qubo_coefficients1}) and (\ref{eq:qubo_coefficients2}), must be provided to the D-Wave Ocean Software \cite{dwave_ocean}, the interface that communicates with the quantum hardware. 

The readout of the quantum algorithm is a bit string $q_0q_1...q_{NR-1}$ of $N.R$ qbits that returns to the classical algorithm for further processing. In principle, the qubits must be fully connected. However, in situations where the matrix $A$ is sparse, several coefficients $b_{rs}^{ij}$ are set to zero, hence not all connections between the qubits are necessary. With knowledge of the above, if the full linear system of equations is solved directly, the number of required qubits $N.R$ for equivalent floating-point accuracy of classical algorithms makes this implementation prohibitive at the current stage of technological development. However, the linear system can be divided into subsystems, and the full linear system solved iteratively, in a hybrid quantum-classical approach, as detailed next.

\subsection{\label{sec:Iterative QUBO}Iterative QUBO method for solving linear systems}

The block Gauss-Seidel method consists in solving the linear system $A\mathbf{x} = \mathbf{b}$ iteratively, by applying the Gauss-Seidel method on the original system divided in blocks \cite{matrixcomputations}. Suppose, for simplicity, that the size of the system is an even number $N$ and divide it into two subsystems,
\begin{align}
    \bmqty{A_{11} & A_{12} \\ A_{21} & A_{22}} \bmqty{\mathbf{x}_1 \\ \mathbf{x}_2} = \bmqty{\mathbf{b}_1 \\ \mathbf{b}_2}  \ .
\end{align}
This translates to 
\begin{align}
    A_{11} \mathbf{x}_1 + A_{12} \mathbf{x}_2 &= \mathbf{b}_1, \ \label{eq:ls_blocks2x2_1}\\
     A_{21} \mathbf{x}_1 + A_{22} \mathbf{x}_2 &= \mathbf{b}_2 \ . \label{eq:ls_blocks2x2_2}
\end{align}

In the iterative method we make an initial guess for $\mathbf{x}_2$, such as $\mathbf{x}_2^{(0)} = 0$, where the upper index of the block-vector $\mathbf{x}_2^{(0)}$ defines the step of the iteration process. Rearrangement of equations (\ref{eq:ls_blocks2x2_1}) and (\ref{eq:ls_blocks2x2_2}) allow us to set up an iterative process where the initial guess is updated by solving two linear systems of size $N/2$. 
\begin{align}
        A_{11} \mathbf{x}_1^{(1)}  &= \mathbf{b}_1 - A_{12} \mathbf{x}_2^{(0)} \ , \label{eq:ls_it_1_blocks2x2_1}\\
      A_{22} \mathbf{x}_2^{(1)} &= \mathbf{b}_2 - A_{21} \mathbf{x}_1^{(1)}\ . \label{eq:ls_it_1_blocks2x2_2}
\end{align}
At the $k$-th step, 
\begin{align}
     A_{11} \mathbf{x}_1^{(k)}  &= \mathbf{b}_1 - A_{12} \mathbf{x}_2^{(k-1)} \ , \label{eq:ls_it_p_blocks2x2_1} \\
     A_{22} \mathbf{x}_2^{(k)} &= \mathbf{b}_2 - A_{21} \mathbf{x}_1^{(k)} \ . \label{eq:ls_it_p_blocks2x2_2}
\end{align}
If the sequences $\mathbf{x}_1^{(k)}$ and $\mathbf{x}_2^{(k)}$ converge, then $\mathbf{x}_1^{(k)} \simeq \mathbf{x}_1^{(k-1)}$, $\mathbf{x}_2^{(k)} \simeq \mathbf{x}_2^{(k-1)}$ and the original linear system is satisfied. Supposing $A_{11}$ and $A_{22}$ are invertible, it can be shown that,
\begin{align}
        \mathbf{x}_1^{(k)} - \mathbf{x}_1 &= \left(A_{22}^{\, -1} A_{21} A_{11}^{\, -1} A_{12}\right)^k  \left( \mathbf{x}_1^{(0)} - \mathbf{x}_1 \right) \ , \\
        \mathbf{x}_2^{(k)} - \mathbf{x}_2 &= \left(A_{11}^{\, -1} A_{12} A_{22}^{\, -1} A_{21}\right)^k \left( \mathbf{x}_2^{(0)} - \mathbf{x}_2 \right) \ .
\end{align}
Hence, a sufficient condition for the convergence of this method is: 
\begin{align}
        \norm{A_{22}^{\, -1} A_{21} A_{11}^{\, -1} A_{12}} &< 1 \ , \\
        \norm{ A_{11}^{\, -1} A_{12} A_{22}^{\, -1} A_{21}} &< 1 \ .
\end{align}

The generalization of the method is possible by dividing the original linear system into several blocks whose size is adjustable to the particular interest. If the system is divided into $D$ blocks, each smaller system has size $N/D \times N/D$. The number of required iterations is not known beforehand, but defined by a convergence criterion. A typical criterion depends on the normalized residual $r^{(k)}$, defined by
\begin{align} \label{eq:residual}
 r^{(k)} \equiv \frac{\norm{A\mathbf{x}^{(k)} - \mathbf{b}}}{\norm{\mathbf{b}} } \end{align}
where $\mathbf{x}^{(k)}$ is the solution of the system of linear equations after $k$ iteration steps. A solution is considered converged when $r^{(k)}$ meets a specified tolerance $\tau$, such that $r^{(k)} \leq\tau$. Here, we instead monitor the relative error of the solution vector at the $k$-th iteration $\mathbf{x}^{(k)}$, given as,
\begin{align} \label{eq:error}
e^{(k)}\equiv\frac{\norm{\mathbf{x}^{(k)}-\mathbf{x}}}{\norm{\mathbf{x}}}\, .  
\end{align}
It can be shown that $e^{(k)}$ and $r^{(k)}$ are related according to,
\begin{align}
e^{(k)}\leq \kappa (A) r^{(k)} \, ,
\end{align}
where $\kappa (A)$ is the condition number of the coefficient matrix $A$. Preconditioning techniques should be used when $A$ is ill-conditioned. In this work we do not pursue this issue further, since it would be relevant for both classical and quantum algorithms.

\subsubsection*{\label{sec:scaling}Scaling of the method}


The representation of a vector $\mathbf{x}$ with $N$ components requires $N \cdot R$ qubits, where $R$ is the number of qubits that gives the desired accuracy of floating-point numerical representation, $\epsilon$, of each variable $x_i$. However, as stated above, it is possible to solve smaller subsystems of size $N/D$ by partitioning the full system in $D$ parts. In this case the number of qubits scales as $(N/D)\cdot R$. In order to reduce the number of qubits required in the computation for a given floating-point accuracy of $x_i$, one possibility is to shrink the interval $[-d_i, 2c_i - d_i]$ at each iteration, which we adopt here. Any reduction in the number of required qubits is significant in the era of noisy intermediate-scale quantum (NISQ) computers, where at most few thousands of qubits are available. Even more dramatic is the number of connections necessary to solve the QUBO problems. In the worst case, every logical qubits must be fully connected (see the coefficient $b_{rs}^{ij}$ in Eq. (\ref{eq:qubo_coefficients1})). The block Gauss-Seidel method can reduce the number of connections among qubits  by a factor $N^2R^2 \left(1 - 1/D^2 \right)$. Additionally, when $A$ is sparse, connectivity is reduced considerably. 


If $\epsilon$ is a bound for the accuracy of the floating-point numerical representation of a physical variable $x_i$, for a solution represented with $R$ qubits, then
\begin{align*}
\frac{2 c_i}{2^R} \le \epsilon \ ,
\end{align*}
since $2 c_i$ is the length of the open interval $[-d_i, 2c_i - d_i)$ where the solution is represented and $2^R$ is the number of different points that can be represented with $R$ bits inside the aforementioned interval. From the expression above, the number of qubits required to represent one variable with  floating-point accuracy $\epsilon$ scales as
\begin{align*}
R \ge \log_{2} \left( 2c_i/\epsilon\right)\ .
\end{align*}

Since the total computing time $\mathcal{T}$ depends on the energy gap between the ground and first excited states of the system, it is hard to obtain precise estimates of $\mathcal{T}$. As noted in \cite{amin2009}, the task of analytically extracting the minimum gap scaling has been extremely difficult in practice, except for a few special cases, such as for Deutsch-Jozsa \cite{das2002adiabatic} and Grover algorithms \cite{roland2002quantum,farhi2000grover}.

In Table~\ref{tb:qubo_comparison} we summarize the consuming quantum resources used in the iterative and non-iterative QUBO algorithms to solve the linear system of equations.
\begin{table}[h]
\begin{center}
\begin{tabular}{c|c|c}
    Method  &  \ \ \# qubits \ \ & time \\ \hline
    Original QUBO  & $N \cdot R$     &  $\mathcal{T}_{N,R}$ \\
    Iterative QUBO & $  (N/D) \cdot R$ &  $\mathcal{T}_{N/D,R} \cdot D \cdot n_\text{iter}$
\end{tabular}
\caption{Comparison between original QUBO and iterative QUBO methods. $\mathcal{T}_{N,R}$ indicates the time for running one QUBO algorithm for a linear system of size $N$ and $R$ qubits for representing the solution, while $\mathcal{T}_{N/D,R}$ is the time for solving a system of size $N/D$. $n_\text{iter}$ is the number of iterations in the case of the iterative QUBO method. Both times $\mathcal{T}_{N,R}$ and $\mathcal{T}_{N/D,R}$ depend on the annealing time and the number of runs to sampled solutions of the problem.}
\label{tb:qubo_comparison}
\end{center}
\end{table}

%% file: results.tex
\section{\label{sec:results}Results and Discussion: Two Dimensional Heat Equation}

\begin{figure}[htpb]
\begin{center}
\includegraphics[scale=0.6]{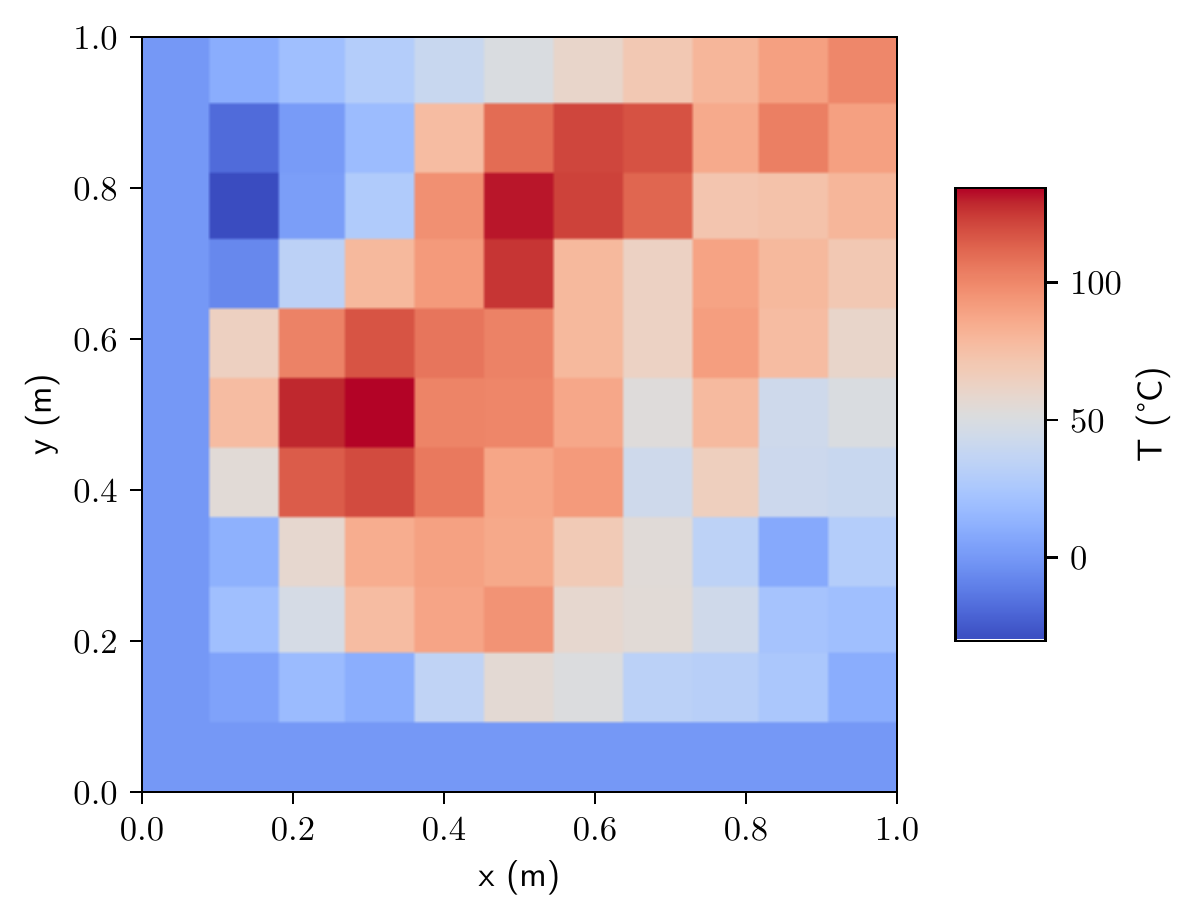}
\caption{\label{fig:HeatMap} 
Temperature distribution in the square plate.}
\end{center}
\end{figure}

\begin{figure}[htpb]
\begin{center}
\includegraphics[scale=0.5]{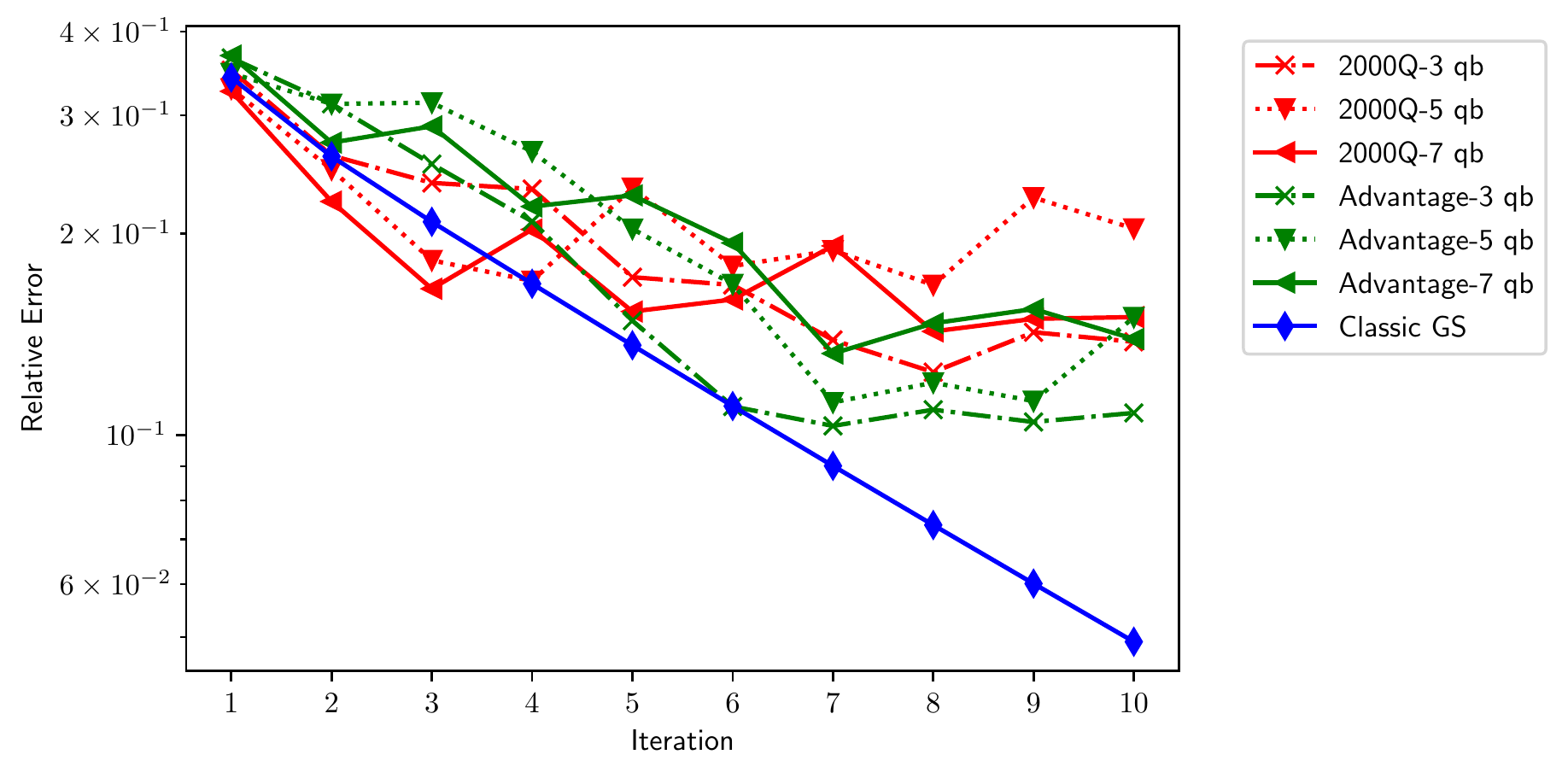}
\caption{\label{fig:Error_vs_Iterations} 
The relative error as a function of the number of iterations of the block Gauss-Seidel method for different numbers of qubits used in the numerical representation of the solution $\mathbf{x}$. The solutions obtained by the 2000Q (red) and Advantage (green) systems are compared with the classic Gauss-Seidel implementation (blue).}
\end{center}
\end{figure}

\begin{figure}[htpb]
\begin{center}
\includegraphics[scale=0.5]{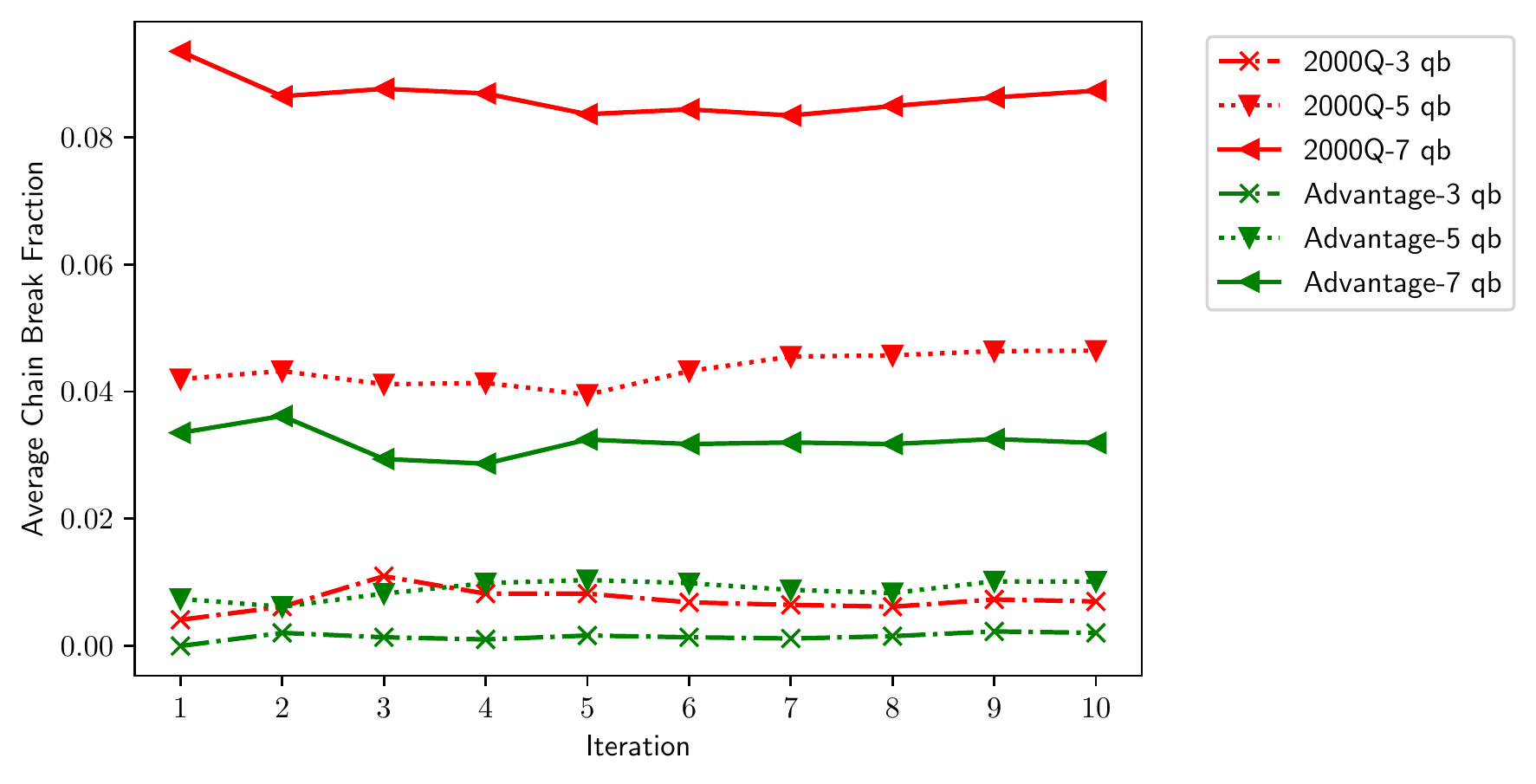}
\caption{\label{fig:ChainBreak_vs_Iterations} 
Average chain break fraction as a function of the number of iterations in the block Gauss-Seidel method for different numbers of precision qubits.}
\end{center}
\end{figure}

\begin{figure}[htpb]
\begin{center}
\includegraphics[scale=0.5]{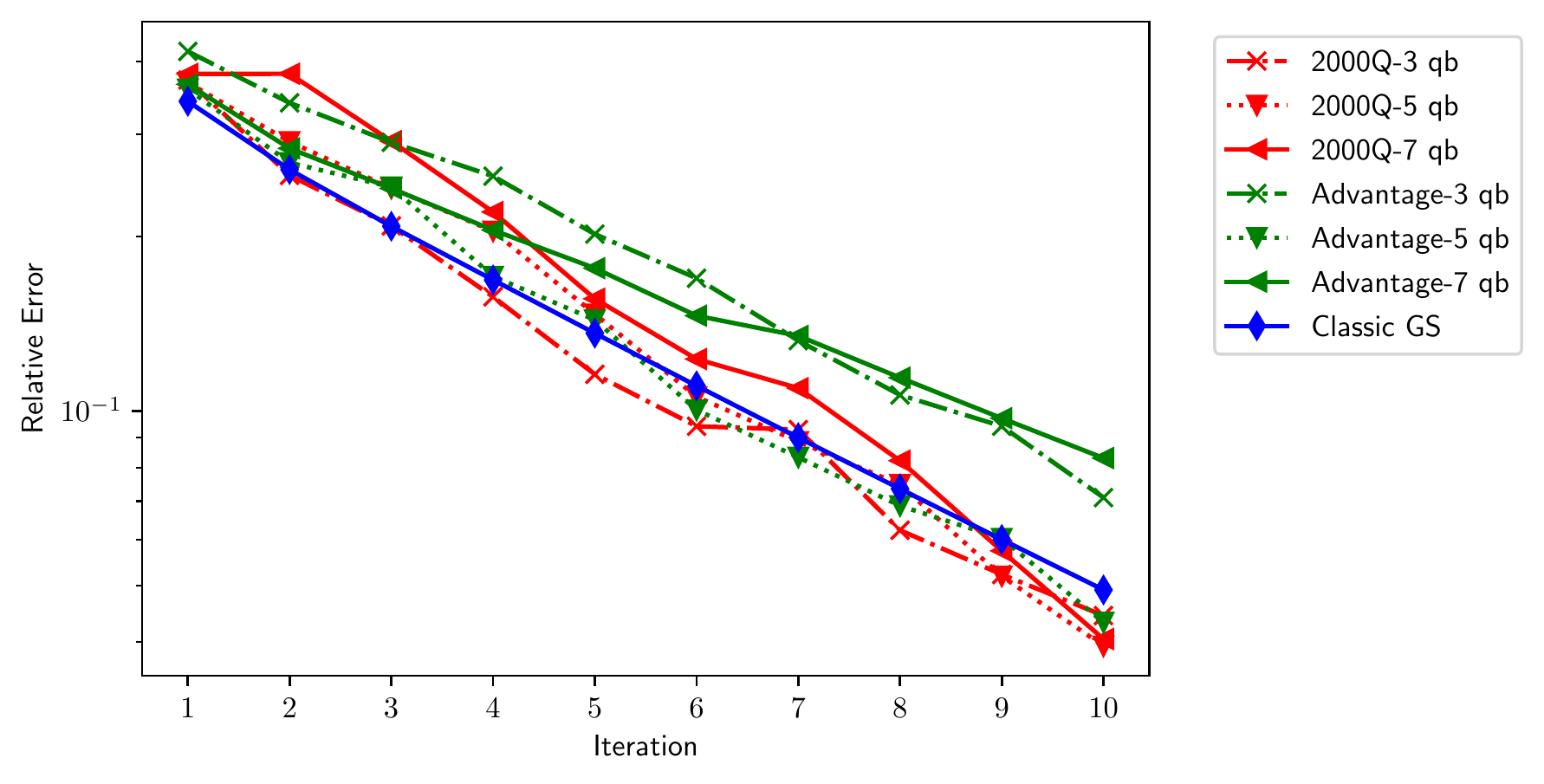}
\caption{\label{fig:Error_vs_Iterations_shrink} The relative error as a function of the number of iterations of the block Gauss-Seidel method, with a shrink factor of $0.8$ of the search domain at each iteration, for different numbers of qubits used in the numerical representation of the solution $\mathbf{x}$. The solutions obtained by the 2000Q (red) and Advantage (green) systems are compared with the classic Gauss-Seidel implementation (blue).}
\end{center}
\end{figure}

In order to assess the viability and performance of the proposed method, we solve the heat equation for a square plate of length $L$ subject to fixed temperature at its edges and multiple heat sources and sinks within the domain. The steady state solution $T(x,y)$ is then determined with the following PDE and boundary conditions,
\begin{align}\label{eq:pde_heat_eq}
    \dfrac{\partial^2 T}{\partial x^2} + \dfrac{\partial^2 T}{\partial y^2} = 0 \,\,\,,
    \end{align} 
    \begin{align}\label{eq:boundary_conditions_heat_eq}
    T(x,0) &=  0^\circ C \ , \ \ \quad \quad \ \ x \in [0,L] \nonumber \\
    T(x,L) &=  \frac{x}{L} \cdot 100^\circ C  \ , \ \ x \in [0,L]   \nonumber\\
    T(0,y) &=  0^\circ C  \ ,  \ \ \quad \quad \ \  y \in [0,L]   \nonumber\\
    T(L,y) &= \frac{y}{L} \cdot 100^\circ C  \ , \ \ y \in [0,L] \ .
\end{align}   

The problem is solved numerically by the finite difference method. The domain is divided into $m$ segments of equal length, with nodes located at positions $x_i = (L/m) \cdot i$ and $y_j = (L/m) \cdot j$, where $i,j=0\dots m$. The points $(x_i,y_j)$ of the grid corresponding to $i=0$, $i=m$, $j=0$ or $j=m$ form the boundary of the plate and are determined by the boundary conditions (\ref{eq:boundary_conditions_heat_eq}). The points inside the grid are the unknowns, and the partial derivatives can be written as
\begin{align}
\begin{split}
        \frac{\partial^2 T}{\partial x^2}(x_i,y_j)  =& \frac{T(x_{i+1},y_j) -2T(x_{i},y_j) + T(x_{i-1},y_j)}{(L/m)^2} \\
        +&\frac{(L/m)^2}{12}\dfrac{\partial^4 T}{\partial x^4}(x_i^{\ast},y_j) \,\,, \\
        \frac{\partial^2 T}{\partial y^2}(x_i,y_j)  =& \frac{T(x_{i},y_{j+1})- 2T(x_{i},y_j) + T(x_{i},y_{j-1})}{(L/m)^2} \\
        +& \frac{(L/m)^2}{12}\dfrac{\partial^4 T}{\partial y^4}(x_i,y_j^{\ast}) \,\,\,,
\end{split}
\end{align}
for some $x_i^{\ast} \in [x_{i-1},x_{i+1}]$ and $y_j^\ast \in [y_{j-1},y_{j+1}]$. Ignoring terms proportional to $(L/m)^2$ or higher orders, the PDE becomes,
\begin{align}\label{eq:pde_heat_eq_ls}
  \begin{split}
    T(x_{i+1},y_j) &- 4T(x_{i},y_j) + T(x_{i-1},y_j) \\ &+ T(x_{i},y_{j+1}) + T(x_{i},y_{j-1}) = 0 \,,
  \end{split}
\end{align}
 where $i,j = 1, 2, \ldots, m-1$. Equation (\ref{eq:pde_heat_eq_ls}) can be written in the form of a linear system $A\textbf{x} = \mathbf{b}$, where $A$ is a sparse matrix with few non-zero entries and $\mathbf{b}$ is defined by the boundary conditions and eventual source terms. 

The spatial discretization forms a $9\times 9$ matrix ($m=11$), resulting in a system of coupled linear equations with $81$ unknown variables, which is solved by the iterative QUBO algorithm on the Advantage and D-Wave 2000Q systems \cite{dwave_docs}. A sample solution of the problem is shown in Fig.~\ref{fig:HeatMap}.

This approach enables us to solve larger linear systems of equations, circumventing the limited number of qubits in the D-Wave quantum processing units. For sake of comparison, in this work we use at most $7$ precision qubits, which demands at most $56$ logical qubits to describe all unknown variables in the block Gauss-Seidel approach, while $567$  would be required to solve the full system of linear equations non-iteratively. We will see that the number of logical qubits is much smaller than the number of physical qubits in both D-Wave systems, because of the limited connectivity between physical qubits.

In Fig.~\ref{fig:Error_vs_Iterations} the dependence of the relative error defined in Eq.~(\ref{eq:error}) is shown as a function of the number of iterations of the block Gauss-Seidel method, for different numbers of qubits used in the numerical representation of the unknown variables. In this example, the smallest error is found using three qubits in the Advantage system. We also notice that the error does not continually decrease, as would be expected in a classical algorithm. This occurs in part because of the limited floating-point accuracy of the variables $x_i$, such that after a certain number of iterations, an improvement of the solution is not possible. Another observation is that the increase in qubits does not necessarily translate into smaller errors. This result was not expected. However, it can be related to the appearance of more noise in the system, since there are more qubits involved in the numerical representation of a given variable $x_i$, which in turn increases the the probability of chain breaks. This subject will be discussed in the following. In general, the Advantage system performs marginally better than 2000Q. 

In order to create a logical qubit, generally more than one physical qubit is necessary. This is a consequence of the limited number of couplers (connections) between qubits. In the D-Wave quantum processing units, a set of physical qubits that describe a vertex of a graph form a chain. For this chain to behave as a basic quantum processing unit, all of its physical qubits must be in the same state. When this does not occur, the chain is broken and the state is determined by a majority vote. In Fig. (\ref{fig:ChainBreak_vs_Iterations}) we plot the average chain break fraction as function of the number of iterations and the number of qubits used to represent each value of temperature. The average chain break fraction for each number of precision qubits is greater in the 2000Q than in the Advantage system and is on average independent on the iteration number. This result is expected, since in the Chimera topology of the 2000Q system each qubit is connected to $6$ other qubits, while in the Pegasus topology of the Advantage system, each qubit connects to $15$ other qubits. Furthermore, the Advantage system has $5000+$ qubits compared to $2000+$ qubits in the 2000Q system. These features allow for more compact chains that are less susceptible to noise \cite{dwave_advantage}.

A significant improvement in the relative error of Fig.~\ref{fig:Error_vs_Iterations} can be obtained by reducing the length of the interval $2c_i$ represented by a given number of qubits at each iteration. The iterative QUBO algorithm is modified by shrinking the interval by factor of $\gamma=0.8$ at each iteration, i.e., the interval for estimating the solution at the $k$-th iteration is given by 
$[x_i^{k-1} - c_i \gamma^{k-1},x_i^{k-1} + c_i\gamma^{k-1}]$, where $x_i^{k-1}$ is the estimate for $x_i$ at the ($k-1$)-th iteration. This procedure increases the capacity of the QUBO solver to represent the solution of the system with a fixed number $R$ of qubits, as can be seen in Figure~\ref{fig:Error_vs_Iterations_shrink}. The implementation of this updated search interval length brings the quantum solvers on par with the error obtained on classical hardware. It must be noted, however, that the update of the rhs of equations~(\ref{eq:ls_it_p_blocks2x2_1}) and~(\ref{eq:ls_it_p_blocks2x2_2})  during the Gauss-Seidel iterations is performed on a classical computer, with double precision, even in the quantum algorithm.

A comparison of Figs. \ref{fig:Error_vs_Iterations} and \ref{fig:Error_vs_Iterations_shrink} shows that the 2000Q performs slightly better than the Advantage system when the update of the search interval length is incorporated into the algorithm. There is not a clear interpretation why this occurs, since the Advantage system is a improved version of the 2000Q system, where the chains are more compact and less susceptible to noise \cite{dwave_advantage}. The results obtained from both figures also suggest that a greater number of precision qubits does not necessarily translate to a smaller error, as already noticed earlier.

%% file: acknowledgments.tex
\begin{acknowledgments}

The authors acknowledge the financial support from the Brazilian funding agencies Coordenação
de Aperfeiçoamento de Pessoal de Nível Superior - CAPES, Conselho
Nacional de Desenvolvimento Científico e Tecnológico - CNPq, Fundação
de Amparo à Pesquisa e Inovação do Estado de Santa
Catarina - FAPESC, and the Instituto Nacional de Ciência e Tecnologia
de Informação Quântica - INCT-IQ.

\end{acknowledgments}

%% file: conclusions.tex
\section{\label{sec:conclusion} Conclusions and perspectives}

The hybrid classical-quantum algorithm we have developed solves partial differential equations by the finite difference method, yielding a system of coupled linear equations. This technique was tested with success for the 2D steady-state heat equation, but in principle can be applied to any partial differential equation, encompassing a large array of applications for solution by quantum annealers. The method with updated search interval length allows us to obtain comparable accuracy to a classical algorithm with double precision using 3 qubits to numerically represent each variable. The main limitation of the quantum hardware is the reduced connectivity between qubits, albeit great strides have been made in order to increase the number of couplers for each qubit. The iterative algorithm allows for a significant reduction in the number of logical qubits, from $N.R$ to $N.R/D$, at the expense of an increase in the number of required iterations for a given accuracy. 

It is our understanding that there are good prospects for the iterative QUBO algorithm and that there is room for further improvements, such as robust heuristics for update of the search interval length. Another improvement can be obtained by using an application-oriented embedding, as many linear systems typical of finite difference and finite volume methods are characterized by sparseness of the coefficient matrix $A$. This has the advantage of requiring less couplers than a fully connected problem. Possibly a specific embedding can be found for this class of problems, allowing for less qubits and enabling the solution of larger systems of coupled linear equations. Owing to our limited access to the D-Wave QPUs, we believe that more tests are required to fully benchmark the performance of 2000Q and Advantage systems for the solution of systems of linear equations.